\documentclass[useAMS,usenatbib]{mn2e}
\usepackage[utf8]{inputenc}
\usepackage{fixltx2e}
\usepackage{color}
\usepackage[usenames,dvipsnames,svgnames,table]{xcolor}
\usepackage{graphicx}
\usepackage{amsmath}

\title[Swift J045106.8-694803; a highly magnetised neutron star in the Large Magellanic Cloud.]{Swift J045106.8-694803; a highly magnetised neutron star in the Large Magellanic Cloud.}
\author[H. Klus, E.S. Bartlett, A.J. Bird, M. Coe, R.H.D. Corbet and A. Udalski]{H. Klus$^{1}$\thanks{E-mail: hvk1g11@soton.ac.uk (HK)}, 
E.S. Bartlett$^{1}$, A.J. Bird$^{1}$, M. Coe$^{1}$, R.H.D. Corbet$^{2}$ and A. Udalski$^{3}$\\
$^{1}$The Faculty of Physical and Applied Sciences, University of Southampton, Highfield, Southampton SO17 1BJ, United Kingdom\\
$^{2}$University of Maryland Baltimore County, X-ray Astrophysics Laboratory, Mail Code 662, NASA Goddard Space Flight Center,\\Greenbelt, MD 20771, USA\\
$^{3}$Warsaw University Observatory, Aleje Ujazdowskie 4, 00-478 Warsaw, Poland}
\begin{document}

\date{Accepted 2012 October 28. Received 2012 October 19; in original form 2012 September 10}

\pagerange{\pageref{firstpage}--\pageref{lastpage}} \pubyear{2012}

\maketitle

\label{firstpage}

\begin{abstract}

We report the analysis of a highly magnetised neutron star in the Large Magellanic Cloud (LMC). The high mass X-ray binary pulsar Swift J045106.8-694803 has been observed with Swift 
X-ray telescope (XRT) in 2008, The Rossi X-ray Timing Explorer (RXTE) in 2011 and the X-ray Multi-Mirror Mission - Newton (XMM-Newton) in 2012. The change in spin period over these four 
years indicates a spin-up rate of $-5.01\pm0.06$ s/yr, amongst the highest observed for an accreting 
pulsar. This spin-up rate can be accounted for using Ghosh and Lamb’s (1979) accretion theory assuming it has a magnetic field of (1.2$\pm^{0.2}_{0.7}$)$\times$10\textsuperscript{14} Gauss. This is 
over the quantum critical field value. There are very few accreting pulsars with such high surface magnetic fields and this is the first of which to be discovered in the LMC. 
The large spin-up rate is consistent with Swift Burst Alert Telescope (BAT) observations which show that Swift J045106.8-694803 has had a consistently high X-ray luminosity for at least five years. 
Optical spectra have been used to classify the optical counterpart of Swift J045106.8-694803 as a B0-1 III-V star and a possible orbital period of 21.631$\pm$0.005 days has been found from MACHO optical photometry. 

\end{abstract}


\section{Introduction}
High mass X-ray binaries (HMXB) are binary systems composed of either a neutron star, white dwarf or black hole and an optical companion, either a supergiant star 
– in the case of supergiant X-ray binaries (SGXB) – or a dwarf, subgiant or giant OBe star - in the case of Be/X-ray binaries (BeXB). HMXB can be detected 
by the X-ray, optical and infrared emission they produce.

X-rays are produced as matter is transferred from the optical star to its denser companion. It is accelerated as it moves into the gravitational 
potential well produced by the denser star and - if the denser star is a neutron star - it is then channelled by the star's magnetic field lines to the magnetic poles. 
The accreted matter then rapidly decelerates 
when it reaches the surface and potential energy is converted to heat which energises the plasma, producing X-ray hot spots. 
These can appear as `beams' if they periodically travel past our line of sight as the star rotates. The pulsation period is equated with the rotation period of the neutron 
star's crust and this value changes with time as accreted matter transfers angular momentum to or from the star.

HMXB are composed from massive, and therefore relatively short lived, stars and so their presence indicates that a new population of stars formed relatively 
recently \citep{b26}. This may be why more HMXB have been discovered in the Milky Way and the Small Magellanic Cloud (SMC) than the Large Magellanic Cloud (LMC).

The HMXB in the Milky Way are mostly found within the spiral arms \citep{col} and the SMC underwent a burst 
of star formation about 200 million years ago when it was about three times closer to the LMC than it is today \citep{b14,b15}. The SMC has an HMXB population 
which is comparable in number to the Galaxy despite being almost two hundred times less massive. Whilst about half of the HMXB in the Galaxy are BeXB, all but one HMXB in the 
SMC is a BeXB system. The LMC, on the other hand, contains relatively few HMXB, most of which are BeXBs \citep{b27,b28,b18}.

Swift J045106.8-694803 is part of a newly discovered BeXB system located in the LMC. It was first discovered by Beardmore \emph{et al} (2009) at an RA, Dec (J2000) of 04:51:06.8 and 
-69:48:03.2 respectively, with an uncertainty of 3.5\arcsec. Beardmore \emph{et al} measured an X-ray flux of (1.68$\pm$0.11)$\times$10\textsuperscript{-11} erg 
cm\textsuperscript{-2}s\textsuperscript{-1} over 0.3-10 keV, fit with a power law of photon index 0.96$\pm^{0.06}_{0.04}$ 
with a column density of (1.9$\pm$0.3)$\times$10\textsuperscript{21} cm\textsuperscript{-2}.

The 14-195 keV flux was found to be (2.8$\pm$0.3)$\times$10\textsuperscript{-11} erg cm\textsuperscript{-2}s\textsuperscript{-1} fit with a power law of photon index 2.5$\pm$0.4. 
Beardmore \emph{et al} also report a pulsation period of 187 seconds. The optical companion to this X-ray source is 
the V = 14.70 blue star [M2002] 9775 \citep{b20} with an orbital period of 21.64$\pm$0.02 days.

Grebenev \emph{et al} have recently calculated a photon index of 0.5$\pm$0.5 using data from INTEGRAL. They have also created energy spectra over 3-200 keV showing that the energy of the high 
energy cut-off in the spectrum is at 16.0$\pm$5.0 keV \citep{b23}. 

In this paper we look at Swift J045106.8-694803 in more detail, confirming its luminosity, pulse period, orbital period and spin-up rate which can be used to determine the strength 
of its magnetic field.

\section{Observations}

\subsection{X-ray Observations}
\subsubsection{Swift/XRT}
Swift's X-ray telescope (XRT) is a CCD imaging spectrometer operating over 0.2-10 keV in photon counting mode.
Archival data were taken from NASA's High Energy Astrophysics Science Archive Research Center (HEASARC)\footnote{http://heasarc.gsfc.nasa.gov/} covering 
six time periods from 23rd October 2008 to 3rd October 2011, as summarised in Table 1.

\begin{table}
 \centering
\begin{tabular}{@{}cccc@{}}
  \hline
  Observation ID&Start Date&Start Time&Exposure (ks)\\ 
  \hline
  00038029001&2008-10-23&03:56:38&6.56\\ 
  00038029002&2008-11-11&14:35:46&6.40\\ 
  00038029003&2008-11-14&00:18:15&2.59\\
  00038029004&2009-01-10&13:23:59&4.32\\
  00038029005&2011-09-30&16:50:35&4.47\\
  00038029006&2011-10-03&18:28:46&5.90\\
  \hline
\end{tabular}
\caption{Summary of XRT datasets.}
\end{table}

The images were extracted using the \emph{ftool}\footnote{http://heasarc.nasa.gov/ftools/} \emph{xselect}. Source and background spectra were then extracted from 
regions of 34\arcsec radii. The spectra were binned to have 50 counts per bin. The ancillary response
files (ARF) were calculated with \emph{xrtmkarf} and a redistribution matrix file (RMF) was taken from HEASARC's calibration database (CALDB). 
The position of the source was confirmed using \emph{ftool} \emph{xrtcentroid}.

The total count rate and error of each dataset, as well as the intrinsic H I column density (N\textsubscript{H}) and photon index, were calculated using \emph{ftool} \emph{xspec}. 
The spectra were described by an absorbed power law with a fixed Galactic foreground column density of 8.4$\times$10\textsuperscript{20} cm\textsuperscript{-2} \citep{b21} and 
abundances set in accordance with Wilms, Allen and McCray (2000). Intrinsic absorption and the abundances of elements heavier than helium were set to 0.4 \citep{b33}. X-ray spectra 
were then compiled in \emph{xspec} over 0.2-10 keV.

The 0.2-10 keV flux of each dataset was determined using \emph{xspec}. The luminosity was then calculated using a distance of 50.6$\pm$1.6 kpc \citep{b9}. 

The light curve of each dataset was extracted in \emph{xselect} and a Lomb-Scargle normalised periodogram  was produced for each using time-series analysis package 
\emph{Period}\footnote{http://www.starlink.rl.ac.uk/docs/sun167.htx/sun167.html}. 

\subsubsection{Swift/BAT}
The Swift Burst Alert Telescope (BAT) data were taken with the Hard X-ray Transient Monitor from 16th December 2004 to the 31st May 2010 over 14-195 keV.
A 58 month light curve was downloaded from NASA's Swift/BAT 58-Month Hard X-ray Survey\footnote{http://swift.gsfc.nasa.gov/docs/swift/results/bs58mon/}. 
This contained an average of $\sim$15 observations a day split into 8 energy bands.

\subsubsection{RXTE}
Archival data were taken from HEASARC. 
These were recorded in two datasets on the 28th October 2011 using the Rossi X-ray Timing Explorer's (RXTE) Proportional Counter Array, over $\sim$3-10 keV, as summarised in Table 2.

\begin{table}
 \centering
\begin{tabular}{@{}ccccc@{}}
  \hline
  Observation ID&Start Date&Start Time&Exposure (ks)\\
  \hline
  96441-01-01-00&2011-10-28&13:54:55.6&9.005\\ 
  96441-01-01-01&2011-10-28&10:51:38.6&0.849\\ 
  \hline
\end{tabular}
\caption{Summary of RXTE datasets.}
\end{table}

Cleaned light curves were produced for each dataset and combined. A Lomb-Scargle normalised periodogram was then created using a frequency interval of 
1$\times$10\textsuperscript{-5} Hz.

\subsection{Optical observations}
\subsubsection{Optical photometry}
Archival data were taken from the MAssive Compact Halo Objects (MACHO) project using the 1.27 meter telescope at the Mount Stromlo Observatory in Australia.
This covered the period from 1st November 1992 to the 29th December 1999 and contains instrumental magnitudes using red (R) and blue (B) filters. 

The data were filtered to remove results flagged as erroneous. The four points remaining 
in the R band dataset which were over 2.4 standard deviations from the mean were also removed. This left 96 data points in the red and 207 in the blue. 
A Lomb-Scargle normalised periodogram  was then created in \emph{Period} for both the R and B band datasets using a frequency interval of 
1$\times$10\textsuperscript{-5} Hz and detrending using a polynomial of order 3.

\subsubsection{Optical spectroscopy}
Optical spectra of Swift J045106.8-694803 have been taken on three separate occasions. Red end spectra were taken with the 1.9m Radcliffe telescope at the South 
African Astronomical Observatory (SAAO) on the 12th December 2009 and 26th September 2011. The data were obtained using the unit spectrograph combined with a 1200 lines~mm$^{-1}$ grating 
and the SITe detector at the Cassegrain focus. The resulting spectra have a spectral resolution of $\sim3$~\AA{}. Comparison copper neon spectra were taken immediately 
before and after the observation and were used for wavelength calibration.

The source was also observed on the 8th and 10th December 2011 with the New Technology Telescope (NTT), La Silla, Chile. Grisms 14 and 20 on the ESO Faint Object Spectrograph 
(EFOSC2) were used for blue and red end spectroscopy respectively with a slit width of 1.5\arcsec. Grism 14 has a grating of 600~lines~mm$^{-1}$ that yielded 
1~\AA{}~pixel$^{-1}$ dispersion over a wavelength range of $\lambda\lambda3095$--$5085$~\AA. Grism 20 is one of the two new Volume-Phase Holographic grisms recently 
added to EFOSC2. It has a smaller wavelength range, from 6047--7147~\AA{}, but a superior dispersion of 0.55 \AA{}~pixel$^{-1}$ and 1070 lines~pixel$^{-1}$. Filter 
OG530 was used to block second order effects. The resulting spectra were recorded on a  Loral/Lesser, thinned, AR-coated, UV flooded, MPP CCD with 2048$\times$2048 
pixels, at a spectral resolution of $\sim10$~\AA{} and $\sim6$~\AA{} respectively. Wavelength calibration was achieved using comparison spectra of Helium and Argon 
lamps taken through out the observing run with the same instrument configuration.

The data were reduced using the standard packages available in the Image Reduction and Analysis Facility (IRAF). The resulting spectra were normalized to 
remove the continuum and and a redshift correction applied corresponding to the recession velocity of the LMC, taken as -280 km s$^{-1}$ \citep{Paturel02}.

\section{Results}
\subsection{X-ray Results}
\subsubsection{Swift/XRT}
Figure 1 shows the positions calculated for datasets 00038029001 (red) and 2 (blue). The first dataset 
has an RA, Dec (J2000) of 04:51:06.4 and -69:48:02.5 with a two sigma error radius of 3.6\arcsec. The second has an RA, Dec (J2000) of 04:51:07.0 and -69:48:03.1 with a two sigma error radius 
of 3.6\arcsec. These are consistent with the positions calculated by Beardmore \emph{et al} (2009).

\begin{figure}
 \centering
\includegraphics[scale=0.35,angle=0]{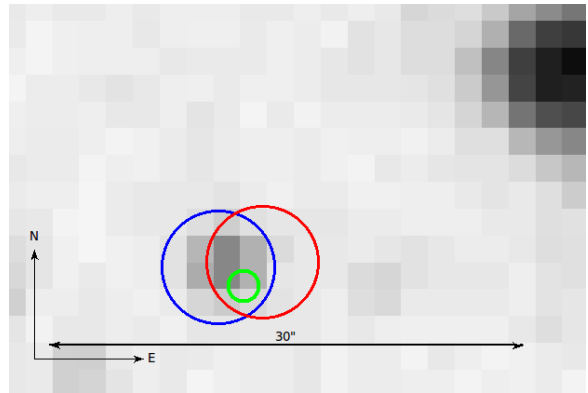}
\caption{Uncertainty circles of the X-ray source overlaid on an optical background (from ESO DSS II Blue). The red circle shows the position found using the Swift/XRT 
dataset 00038029001, the blue circle using dataset 00038029002 and the green using the XMM-Newton dataset from Bartlett \emph{et al} (2012, in prep).}
\end{figure}

The results of data extracted using \emph{xspec} are summarised in Table 3. Figure 2 shows the X-ray spectra of dataset 00038029001 over 0.2-10 keV. This shows that the counts can be 
divided equally into four energy ranges of 0.5-1.5, 1.5-3, 
3-4.5 and 4.5-8 keV. The X-ray luminosities are plotted in Figure 3, which shows that the X-ray luminosities measured between October 2008 and January 2009 were about four 
times higher than those measured in October and September 2011.

\begin{table*}
 \begin{minipage}{140mm}
\begin{tabular}{@{}ccccccccc@{}}           
  \hline
  &Observation ID&Count Rate&N\textsubscript{H}&Photon&Flux&Luminosity&Reduced&Degrees of\\
  &&(counts/s&(10\textsuperscript{21} cm\textsuperscript{-2})&Index&(10\textsuperscript{-11} ergs/&(10\textsuperscript{36} ergs/s)&Chi-Squared&freedom\\
  &&over 0.2-10 keV)&&&s/cm\textsuperscript{2})&&&\\
  \hline
 a&00038029001&0.198$\pm$0.006&3$\pm$2&0.7$\pm$0.1&2.3$\pm$0.1&7.0$\pm$0.3&1.4&21\\ 
 b&00038029002&0.175$\pm$0.005&1$\pm$2&0.7$\pm$0.1&2.1$\pm$0.1&6.3$\pm$0.4&1.2&17\\ 
 c&00038029003&0.182$\pm$0.008&3$\pm$7&0.6$\pm$0.2&2.3$\pm$0.2&7.0$\pm$0.6&0.6&5\\ 
 d&00038029004&0.181$\pm$0.006&1$\pm$3&0.6$\pm$0.1&2.1$\pm$0.1&6.5$\pm$0.4&1.2&11\\
 e&00038029005&0.073$\pm$0.004&15$\pm$14&1.1$\pm$0.4&0.7$\pm$0.2&2.2$\pm$0.5&0.8&2\\
 f&00038029006&0.058$\pm$0.003&1$\pm$16&0.6$\pm$0.4&0.8$\pm$0.2&2.3$\pm$0.5&2.0&2\\
  \hline
\end{tabular}
\end{minipage}
\caption{Summary of information extracted from the spectra of Swift/XRT datasets 00038029001-6 over 0.2-10 keV using a distance of 50.6$\pm$1.6 kpc \citep{b9}.}
\end{table*}

\begin{figure}
\centering
\includegraphics[scale=0.35,angle=270]{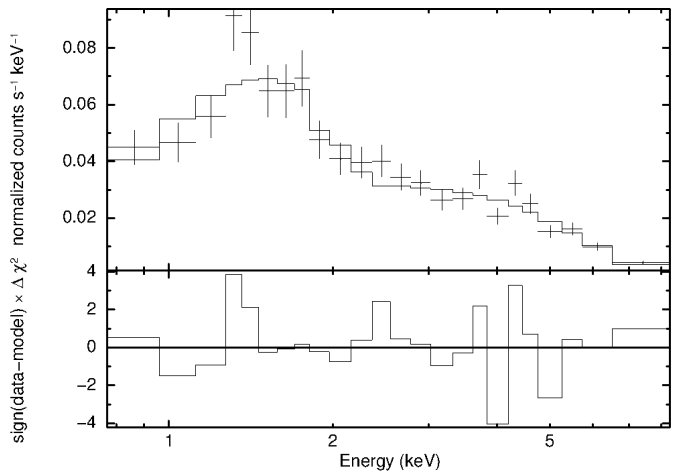}
\caption{The upper panel shows the X-ray spectrum from Swift/XRT dataset 00038029001 over 0.2-10 keV, here the photon index is 0.7$\pm$0.1 with a N\textsubscript{H} of 3$\pm$2 counts/s, 
the lower panel shows residuals.}
\end{figure}

\begin{figure}
 \centering
\includegraphics[scale=0.35,angle=90]{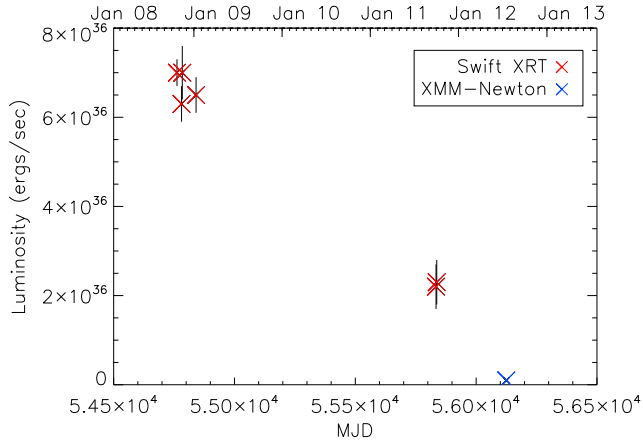}
\caption{Luminosity plotted against time for the Swift/XRT and XMM-Newton datasets over 0.2-10 keV. XMM-Newton dataset from Bartlett \emph{et al} (2012, in prep).}
\end{figure}

Figure 4 shows the Lomb-Scargle normalised periodogram for each of the six datasets. A period is only evident in datasets 00038029001 and 2, these are 187.0$\pm$0.3 and 186.8$\pm$0.2 seconds respectively, 
and show no harmonics. This indicates that light curves folded at these values should be sinusoidal. Folded light curves 
for datasets 00038029001-2 are shown in Figure 5.

Light curves for dataset 00038029001 in each of the four energy ranges determined from Figure 2 are shown in Figure 6. A 
plot of pulse-fraction against energy is shown in Figure 7. The pulse-fraction appears to increase 
with increasing energy between 0.5 and 4.5 keV and then decrease, however the large error bars make this 
inconclusive.

The side peaks in the periodograms are due to gaps in the observations. This 
was confirmed firstly, by fitting a sine wave with the same period to the data, which gave rise to identical peaks. Secondly, by splitting dataset 00038029001 into 
6 shorter datasets composed of continuous observations and adding the individual periodograms, as can be seen in Figure 8.

\begin{figure}   
\centering
\includegraphics[scale=0.35,angle=90]{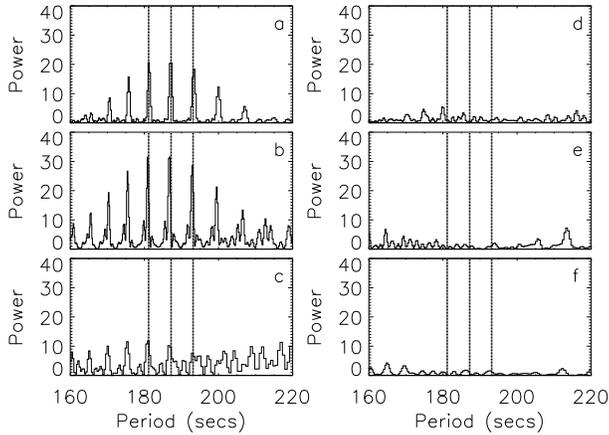}
\caption{Lomb-Scargle normalised periodograms for Swift/XRT datasets 00038029001-6 (labeled a-f) over 0.2-10 keV. The dotted lines indicate the 187 second period 
and the two side peaks of 181 and 193 seconds mentioned by Beardmore \emph{et al} (2009).}
\end{figure}

\begin{figure}
\centering
\includegraphics[scale=0.35,angle=90]{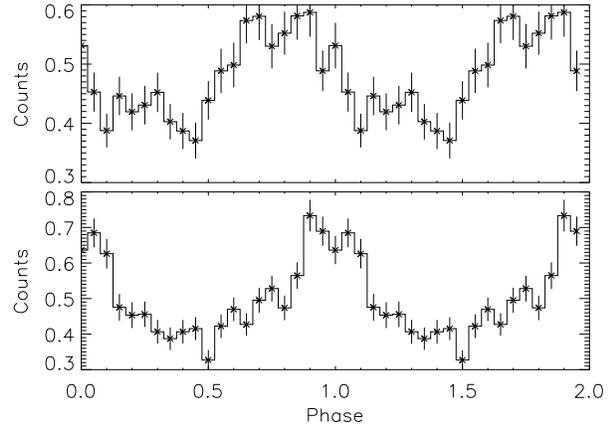}
\caption{Light curves from Swift/XRT datasets 00038029001 (top panel) and 00038029002 (bottom panel) over 0.2-10 keV, folded at 187.0 and 186.8 seconds respectively. The phase shift is arbitrary.}
\end{figure}

\begin{figure}
\centering
\includegraphics[scale=0.35,angle=90]{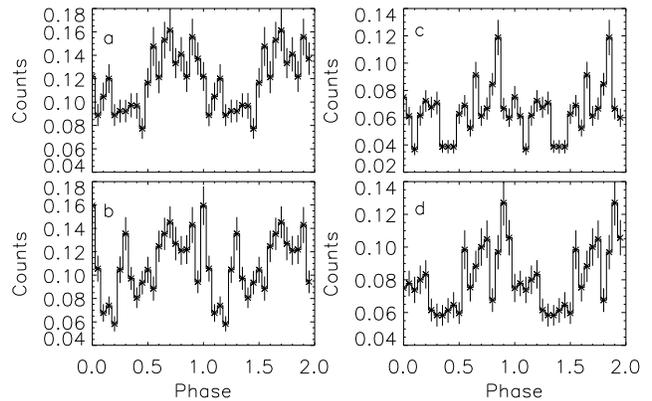}
\caption{Light curves from Swift/XRT dataset 00038029001, folded at 187.0 seconds, over 0.5-1.5, 1.5-3, 
3-4.5 and 4.5-8 keV (labelled a-d).}
\end{figure}

\begin{figure}
\centering
\includegraphics[scale=0.35,angle=90]{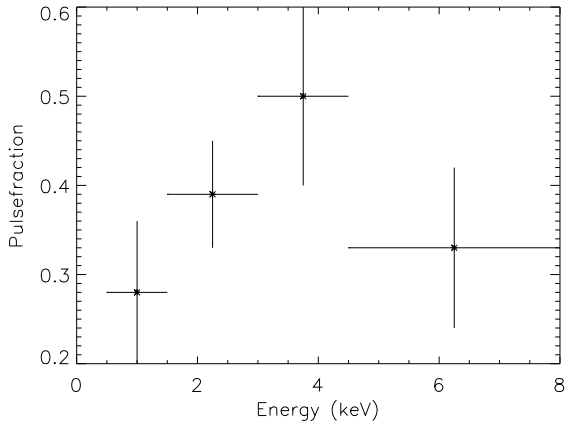}
\caption{Pulse-fraction plotted against energy for Swift/XRT dataset 00038029001 folded at 187.0 seconds, over 0.5-1.5, 1.5-3, 3-4.5 and 4.5-8 keV.}
\end{figure}

\begin{figure}
\centering
\includegraphics[scale=0.35,angle=90]{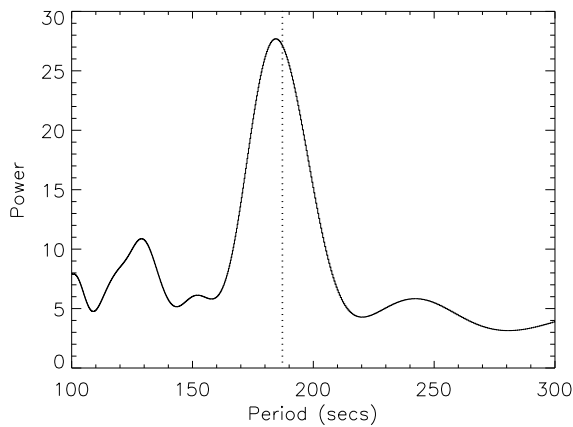}
\caption{Lomb-Scargle normalised periodogram created by splitting dataset 00038029001 into 6 datasets composed of continuous observations and adding the periodograms. 
The dotted line indicates the 187.0 second period found for dataset 00038029001.}
\end{figure}

Datasets 00038029001 and 2 provided the best signal and so their light curves were combined to determine the best estimation for the pulse period in late 2008.
A Lomb-Scargle normalised periodogram yielded a maximum peak at 186.85$\pm$0.04 seconds as can be seen in Figure 9. 
Monte Carlo simulations give this period a 99.9\% confidence level. Bootstrapping was conducted 
in order to confirm that this is the correct peak, and not a product of the window function, and after 5000 iterations between 166 and 200 seconds, a period of 187.07$\pm$0.04
seconds was found 63\% of the time and 181.41$\pm$0.04 seconds 37\% of the time. We therefore concluded that 187.07$\pm$0.04 seconds was the most probable period of Swift J045106.8-694803 
in October/November 2008.

\begin{figure} 
\centering
\includegraphics[scale=0.5,angle=0]{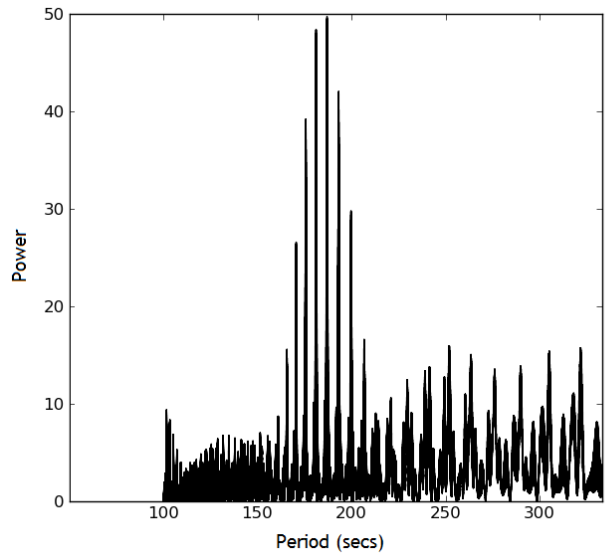}
\caption{Lomb-Scargle normalised periodogram from Swift/XRT datasets 00038029001 and 2 over 0.2-10 keV showing a maximum peak at 186.85$\pm$0.04 seconds. The side peaks are due to gaps in the observations.}
\end{figure}

\subsubsection{Swift/BAT and INTEGRAL/IBIS}
The top panel of Figure 10 shows the long-term light curve of Swift J045106.8-694803 with total counts over 14-195 keV, binned at 28 days. The count rate appears to increase over time, 
peaking in about July 2007 and continuing to remain above zero for the next 
three years. 

This was confirmed by the light curve compiled from the INTEGRAL IBIS data which is shown in the second panel of Figure 10. The bottom panel shows the light curve of the optical counterpart 
taken from 
OGLE III and IV in the I-band. The OGLE data shows that the source was brightest in the I-band whilst it was barely detectable in the X-ray. There is a slight increase in brightness 
after MJD 54000 but this is not significant. 

\begin{figure*}
\centering
\includegraphics[scale=0.65,angle=90]{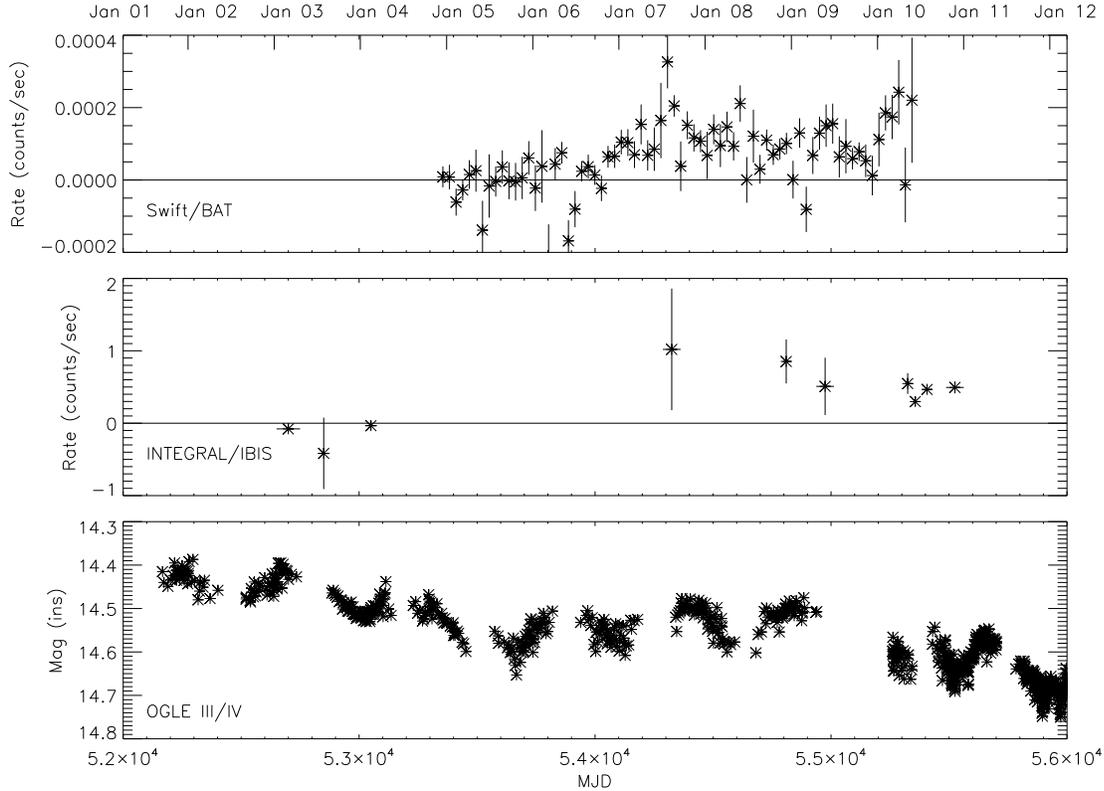}
\caption{Long-term light curve from Swift/BAT (14-195 keV), INTEGRAL IBIS (15 keV - 10 MeV) and OGLE III and IV.}
\end{figure*}

A Lomb-Scargle normalised periodogram  
was created for the BAT dataset using \emph{Period} with a frequency interval of 1$\times$10\textsuperscript{-5} Hz but no periods were detected.

\subsubsection{RXTE}
The Lomb-Scargle normalised periodogram is shown in plots (a) and (b) of Figure 11. As with the Swift/XRT data, the side peaks are due to gaps in the observations. 
This was confirmed by fitting a sine wave with the same period to the data, as can be seen in plot (c) of Figure 11. 
The harmonics at exactly 1/3 and 1/4 of the pulse period, seen in plot (a), indicate that the light curve should be non-sinusoidal and have multiple peaks when folded at the pulse 
period, as shown in Figure 12. 

\begin{figure}
\centering
\includegraphics[scale=0.35,angle=90]{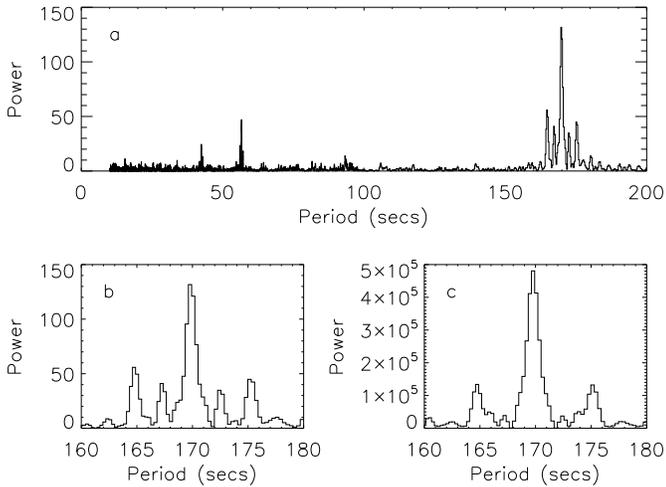}
\caption{Panel (a) shows the Lomb-Scargle normalised periodogram from the combined RXTE datasets over 3-10 keV. Panel (b) shows a close up view of the region 
around the main peak compared to panel (c) which shows the result of a simulated dataset produced by a pure sine wave of period 169.78 seconds.}
\end{figure}

\begin{figure}
\centering
\includegraphics[scale=0.35,angle=90]{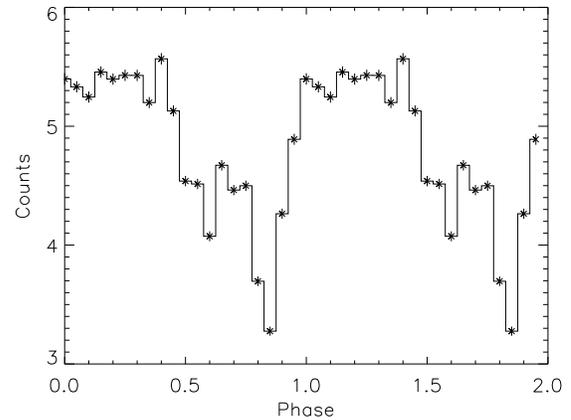}
\caption{Light curve folded at 169.78 seconds from RXTE datasets 96441-01-01-00 and 01 over 3-10 keV.}
\end{figure}

The shape of the light curve contains information on the emission geometry from the regions close to the neutron stars magnetic poles. 
At high luminosities photons can escape from the sides of the accretion column, giving rise to a fan-beam pattern and a more complex profile than that found at lower luminosities 
where X-rays are generally emitted in a pencil-beam.

The RXTE data show a pulsation period of 169.8$\pm$0.3 seconds, 
17.3$\pm$0.3 seconds less than the period evident in the Swift/XRT datasets taken in October and November 2008. 
This gives a spin-up rate ($\dot{P}$) of -5.8$\pm$0.1 s/yr, or (-1.84$\pm$0.03)$\times$10\textsuperscript{-7} s/s, and a spin-up 
time scale (T\textsubscript{s}) of -30.8$\pm$0.5 yr using
\begin{equation}
  T_s=\frac{P}{-\dot{P}}
\end{equation}
 where
$P$ is the average pulse period in seconds.
Since this is an average over three years, it is possible that the spin-up rate has been much higher at points during this period.

\subsubsection{XMM-Newton}
Swift J045106.8-694803 was observed by the X-ray Multi-Mirror Mission - Newton (XMM-Newton) on 17th July 2012 \citep{Bartlett}.
This confirmed the position of Swift J045106.8-694803 - as is shown in Figure 1 - and gave an up-to-date period and luminosity. 
The position was found to be at an RA, Dec (J2000) of 04:51:06.7 and -69:48:04.2 respectively, with a one sigma uncertainty of 1\arcsec. 
The luminosity was found to be (9.8$\pm$0.9)$\times$10\textsuperscript{34} ergs/s - as shown in Figure 3 - and the period 168.5$\pm$0.2 seconds. 

This is 1.3$\pm$0.4 seconds less than the period calculated from RXTE in 2011, giving a $\dot{P}$ of -1.8$\pm$0.5 s/yr. 
It is also 18.6$\pm$0.2 seconds less than the period calculated from Swift/XRT in 2008, giving an average $\dot{P}$ of -5.01$\pm$0.06 s/yr and 
a T\textsubscript{s} of -34.9$\pm$0.4 yr. Figure 13 shows the three periods measured by Swift/XRT in 2008, RXTE in 2011 
and XMM-Newton in 2012. The luminosity and rate of change of spin period may be decreasing but Swift J045106.8-694803 is still continuing to spin-up at a high rate.

\begin{figure}
\centering
\includegraphics[scale=0.35,angle=90]{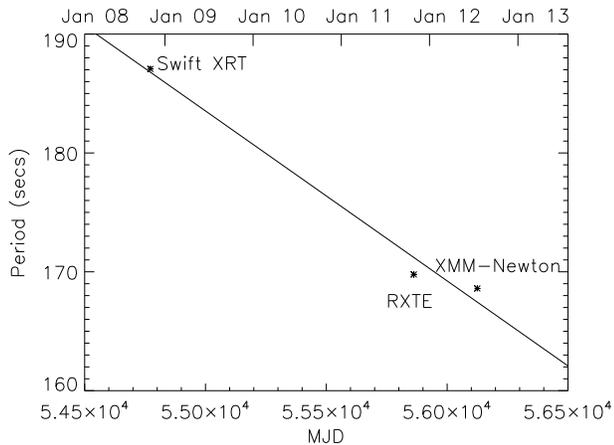}
\caption{Plot showing the three periods measured by Swift/XRT in 2008, RXTE in 2011 and XMM-Newton in 2012 (Bartlett \emph{et al}, 2012, in prep). The straight line is a line of best fit showing that the period is 
continuing to decrease, despite the fact that the rate of change is also decreasing.}
\end{figure}

\subsection{Optical results}
\subsubsection{Optical photometry}
\begin{figure}
\centering
\includegraphics[scale=0.35,angle=90]{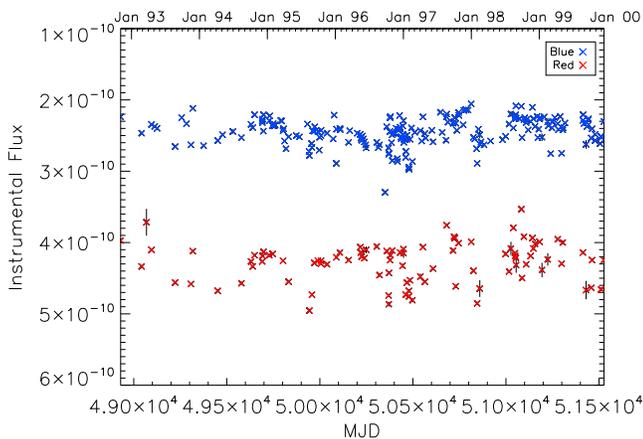}
\caption{Light curves for the R and B band MACHO datasets.}
\end{figure}
Figure 14 shows that the flux of the optical companion to Swift J045106.8-694803 appears to have remained fairly consistent in the B and R bands over 7 years. 
The B band dataset show a possible orbital period of 21.631$\pm$0.005 days as seen in Figure 15, although any underlying non-radial pulsations from the Be star 
may affect the results \citep{b34}. Non-radial pulsations occur when some parts of the stellar surface move inwards while others move outwards at the same time. 
It has been suggested that they could help form the Be stars circumstellar disk. If the rotational velocity of the Be star is close to the critical velocity, then 
pulsations lead to matter being ejected and spun up to form a Keplerian disk \citep{b1}. 

This period is not evident in the R band, this is most likely due to the 
lack of data points rather than the consequence of a real effect. This idea was confirmed by randomly removing half of the B band data points and creating a new Lomb-Scargle normalised 
periodogram, which also failed to show any evidence of an orbital period. A light curve folded at 21.631 days was then produced for the B band dataset as shown in Figure 16. Colour ratio 
and colour magnitude diagrams were also produced. These confirmed that the magnitude in each colour band has remained fairly consistent over the observation period.

\begin{figure}
\centering
\includegraphics[scale=0.35,angle=90]{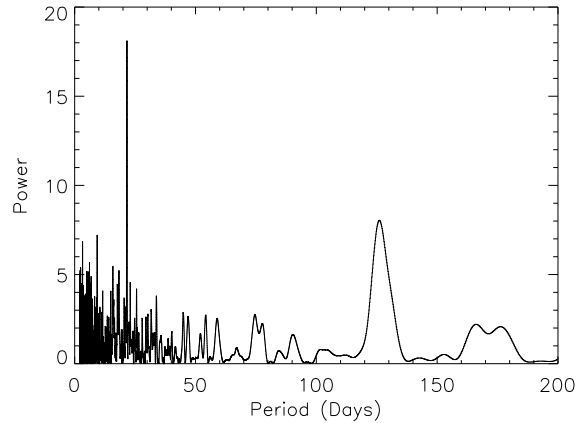}
\caption{Lomb-Scargle normalised periodogram for the B band MACHO dataset showing a possible orbital period of 21.631$\pm$0.005 days.}
\end{figure}

\begin{figure} 
\centering
\includegraphics[scale=0.35,angle=90]{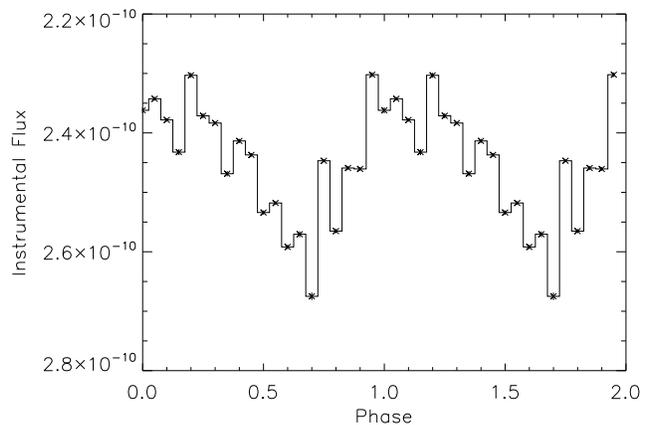}
\caption{Light curve from the B band MACHO dataset folded at 21.631 days.}
\end{figure}

Swift J045106.8-694803 was also observed as part of the OGLE III (LMC136.6.14874) and OGLE IV (LMC531.05.4251) programmes. These I-band data cover, in total, 
a duration of more than a decade, and are shown in the bottom panel of Figure 10. From this figure it is clear that the source undergoes a significant long-term modulation on 
periods in excess of 400 days. Formally, a Lomb-Scargle power spectrum gives the peak to be around 440 days, but a visual inspection of the light curve shows that 
there are other time-scale changes occurring. It is therefore unlikely that this long period is related directly to the binary period, but rather may either 
indicate a general time-scale for fluctuations in the stellar wind, or precessional motion of the circumstellar disk. All these variations make it very difficult 
to search for confirmation of the 21.631 day period seen in the MACHO data. However, if only the better-sampled OGLE IV data are merged with the MACHO data 
(normalised to the approximate starting magnitude of the OGLE III data), then the strength of the 21.631 day peak in the Lomb-Scargle power spectrum increases 
slightly. But without the prior knowledge of this period from the MACHO data such a period would not have been found in the OGLE data.

\subsubsection{Optical spectroscopy}
OB stars in the Milky Way are classified using certain metal and helium line ratios \citep{Walborn90} based on the Morgan-Keenan system (MK; \citealt{MKK1943}). 
However, this is unsuitable in lower metallicity environments as the metal lines are either much weaker or not present. As such, the optical spectrum of IGR~J05414-6858 
was classified using the method developed by \citet{Lennon97} for B-type stars in the SMC and implemented for the SMC, LMC and Galaxy by \citet{Evans04,Evans07}.

Figure 17 shows the unsmoothed optical spectrum of Swift J045106.8-694803. The spectrum is dominated by the hydrogen Balmer series and neutral helium lines. 
There does appear to be evidence for the He\textsc{ii} $\lambda4200$\AA{} line, but it is difficult to distinguish above the noise level along with the He\textsc{ii} 
$\lambda\lambda4541, 4686$\AA{} lines. The He\textsc{i} $\lambda4143$\AA{} line is clearly stronger than the He\textsc{ii} $\lambda4200$\AA{} line constraining the optical 
counterpart of Swift J045106.8-694803 to be later than type O9. There also appears to be evidence for the Si\textsc{iv} $\lambda4088, \lambda4116$\AA{} lines necessary for a 
B1 classification. This is supported by the relative strengths of the Si\textsc{iii} $\lambda4553$\AA{} and Mg\textsc{ii} $\lambda4481$\AA{} suggesting a classification of B2 or earlier.
\begin{figure*}
 \centering
  \includegraphics[height=170mm,angle=90]{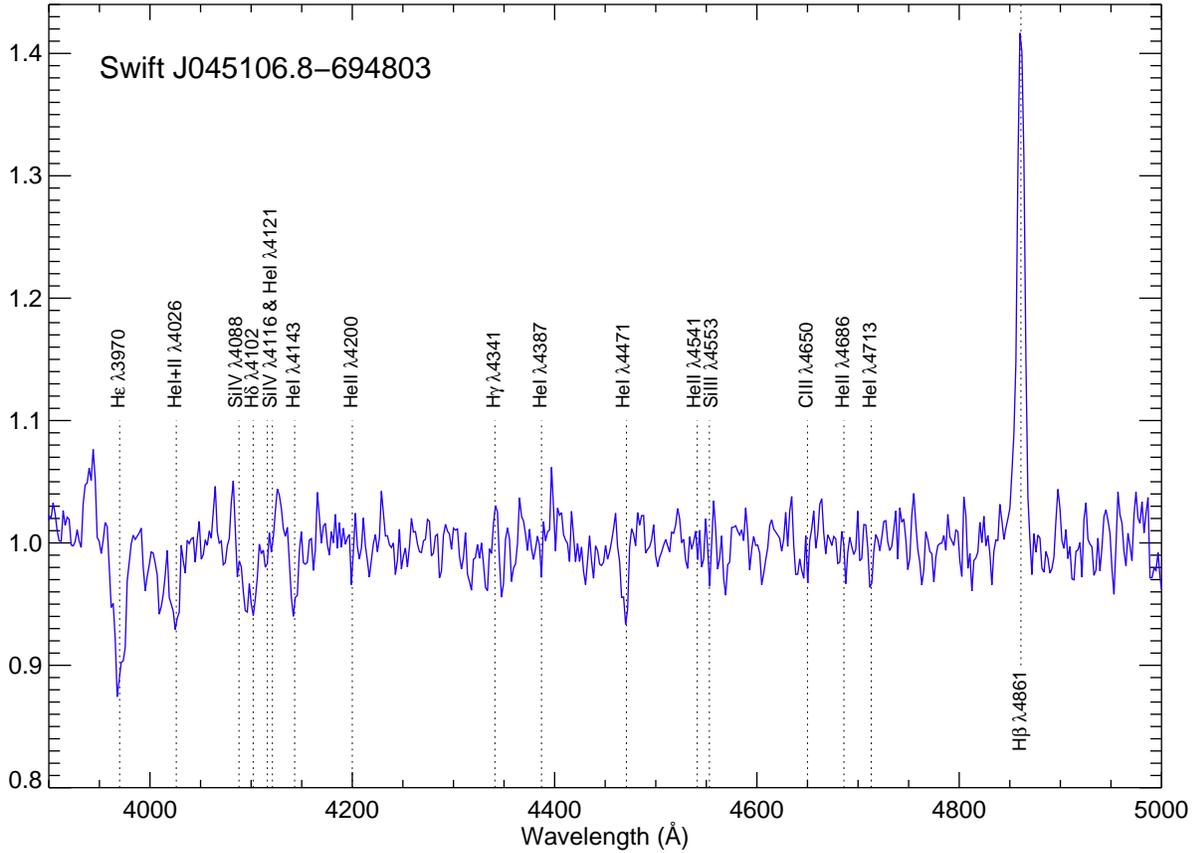}
  \caption{Spectrum of Swift J045106.8-694803 in the wavelength range $\lambda\lambda$3900--5000\AA{} with the NTT on 2011-12-08. The spectrum has been normalized to 
remove the continuum and redshift corrected by -280~km~s$^{-1}$. Atomic transitions relevant to spectral classification have been marked.}\label{fig:blue}
\end{figure*}

The luminosity class of the companion star was determined by the ratios of S\textsc{iv} $\lambda4088$/He\textsc{i} $\lambda\lambda4026-4121$\AA{}, S\textsc{iv} 
$\lambda4116$/He\textsc{i} $\lambda4121$ and He\textsc{ii} $\lambda4686$/He\textsc{i} $\lambda4713$. The first two ratios strengthen with decreasing luminosity class 
(i.e. with increasing luminosity) whereas the latter ratio decreases with increasing luminosity. The relative strengths of these lines are contradictory: The He\textsc{ii} 
$\lambda4686$/He\textsc{i} $\lambda4713$ and S\textsc{iv} $\lambda4116$/He\textsc{i} $\lambda4121$ ratios suggest a luminosity class III, although the proximity of the rotationally 
broadened H$\delta$ $\lambda4102$\AA{} to the Si\textsc{iv} lines makes this more complex. The S\textsc{iv} $\lambda4088$/He\textsc{i} $\lambda4026$ ratio is more consistent 
with a star of luminosity class V. The \emph{V} band magnitude of this star is reported by several sources as between 14.6 and 14.7 (e.g. \citealt{b20, Zaritsky04}). 
An \emph{M$_V$} of -4.3$\pm$0.1 was calculated using a distance modulus of 18.52$\pm$0.07 \citep{b9} along with an \emph{m$_V$} of 14.65$\pm$0.05 and an extinction \emph{A$_V$}=0.4$\pm$0.1 
(calculated using the Galactic column density towards 
the source, $8.4\pm1.0\times10^{20}$~cm$^{-2}$ and the results of \citealt{Guver09}). This is consistent with B0.5III 
\citep{Wegner06}. Less information is available for the absolute magnitude of emission line stars, which will be dependent on the inclination and size of the disc. As such we 
classify the optical counterpart of Swift J045106.8-694803 as a B0-1 III-V star.

\begin{figure}
\centering
 \includegraphics[width=0.45\textwidth]{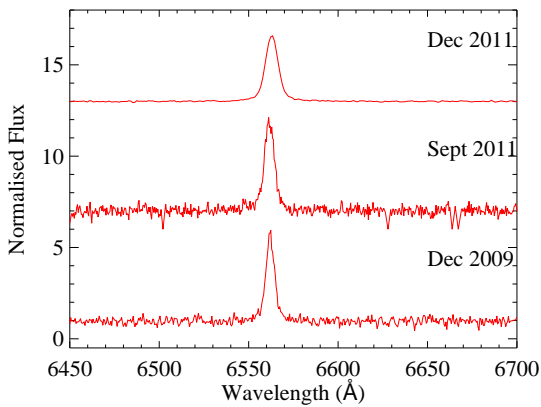}
\caption{ESO (top panel) and SAAO (middle and bottom panels) spectra of Swift J045106.8-694803 in the wavelength range $\lambda\lambda$6400--6700\AA{} with the NTT on 2011-12-10. Spectra have been normalized 
to remove the continuum and shifted by -280~km~s$^{-1}$}\label{fig:red}
\end{figure}
Figure 18 shows the three red end spectra of Swift J045106.8-694803 taken two years apart. The ESO (top) spectrum is offset by 6 flux units. The H$\alpha$ equivalent width, 
considered an indicator for circumstellar disk size, is remarkably similar in all three spectra increasing from -29$\pm$2~\AA{} for the SAAO spectrum taken in 2009 to -33$\pm$1~\AA{} 
and -34.5$\pm$0.6~\AA{} for the SAAO and ESO spectra taken in 2011. It is not uncommon to see large 
variations in the H$\alpha$ equivalent width of these systems on timescales of months. This apparent consistency in the equivalent width (and hence the disk size) is almost certainly 
linked to the exceptionally persistent X-ray activity of the source.

\section{Discussion and conclusions}
We have confirmed the pulsation and orbital periods given by Beardmore \emph{et al} (2009). 
Recent measurements show that Swift J045106.8-694803 has spun-up at a rate of $\sim$5 seconds a year for the last four years and although this rate is decreasing with luminosity, it is still continuing 
to spin-up at a high rate. 

Swift J045106.8-694803 may have experienced one of the highest spin-up rates of any known accreting pulsar but this can be accounted for using Ghosh and 
Lamb’s accretion theory (1979) assuming that the stellar magnetic field is dipolar and that it is accreting from a Keplerian disk. 

\begin{figure*} 
\centering
\includegraphics[scale=0.4,angle=0]{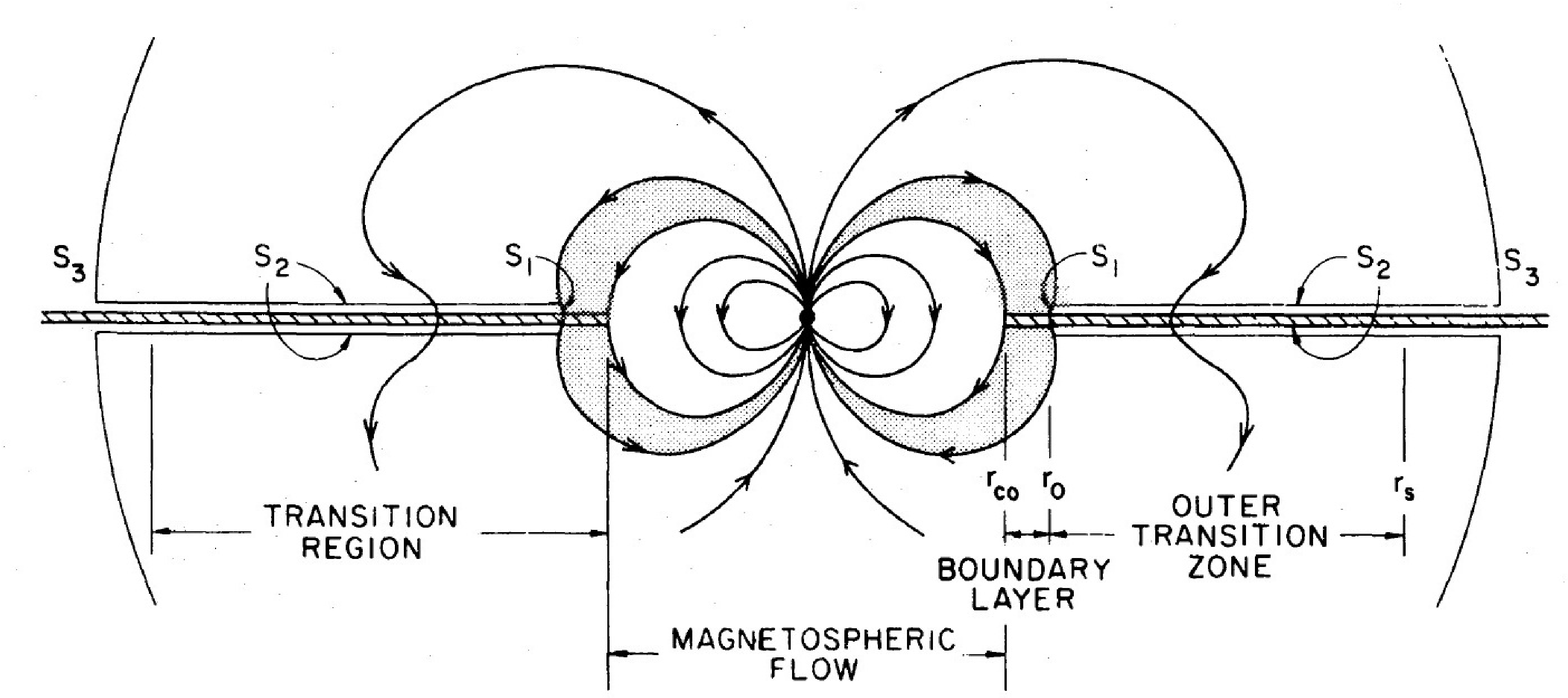}
\caption{Side view of accretion flow from Ghosh and Lamb (1979).}
\end{figure*}

Accretion theory shows that the magnetic field (B) of a disc-fed accreting pulsar is related to the rate of change of its spin period ($\dot{P}$), its average spin period (P) and its average luminosity (L). 

In general, the higher the mass accretion rate ($\dot{M}$), the higher the luminosity and the more angular momentum is accreted per second. 
Where magnetic stresses dominate matter flow at surface S1 – as seen in Figure 19 from Ghosh and Lamb (1979) - this almost always causes spin-up. 

At surface S2, where viscous stresses dominate, this can add a spin-up or spin-down torque depending on the stars fastness parameter ($\omega\textsubscript{s}$). This is the ratio between 
the angular velocity of the neutron star and the Keplerian angular velocity of the disc. 

If the pulse period is long, plasma is able to penetrate into the magnetosphere and reach the neutron star 
surface, transferring angular momentum and causing a strong spin-up torque. If the pulse period is short, the plasma is unable to penetrate further and the neutron star spins down. 

The magnetic moment ($\mu$) can be found if $\dot{P}$, P and L are known by assuming a given mass and radius and seeing which value of $\mu$ predicts the observed $\dot{P}$ using
\begin{equation}
  -\dot{P}=5.0\times10^{-5} \mu_{30}^{2/7} n(\omega_{s}) S_{1}(M) (PL_{37}^{3/7})^2
\end{equation}
where 
$\mu_{30}$ is the magnetic moment in units of 10\textsuperscript{30} Gauss cm\textsuperscript{-3} and 
$L_{37}$ is the luminosity of the accreting star in units of 10\textsuperscript{37} ergs/s. For an 0$<\omega\textsubscript{s}<$0.9
\begin{equation} 
  n(\omega_{s})=1.4(1-2.86\omega_{s})(1-\omega_{s})^{-1} 
\end{equation}
within 5\% accuracy and 
\begin{equation}
  \omega_s={1.35}\mu_{30}^{6/7}S_{2}(M)(PL_{37}^{3/7})^{-1}=1.19P^{-1}\dot{M}_{17}^{-3/7}\mu_{30}^{6/7}(\frac{M}{M_{\odot}})^{-5/7}
\end{equation}
where
$\dot{M}_{17}$ is the mass accretion rate in units of 10\textsuperscript{17} grams/s and
$\frac{M}{M\textsubscript{$\odot$}}$ is assumed throughout to be 1.4.
$S\textsubscript{1}(M)$ and $S\textsubscript{2}(M)$ are structure functions that depends on the mass equation of state and the dynamical response of the neutron star. 
\begin{equation} 
  S_1(M)=R_{6}^{6/7}(\frac{M}{M_\odot})^{-3/7} I_{45}^{-1}
\end{equation}
where
$R\textsubscript{6}$ is the radius of the accreting star in units of 10\textsuperscript{6}cm, assumed throughout to be 1, and 
$I\textsubscript{45}$ is the moment of inertia in units of 10\textsuperscript{45} grams cm\textsuperscript{2}.
\begin{equation} 
  S_2(M)=R_{6}^{-3/7}(\frac{M}{M_\odot})^{-2/7}
\end{equation}
The value of the magnetic field can then be calculated using 
\begin{equation} 
B=\frac{\mu}{R^3}
\end{equation}

Since n($\omega\textsubscript{s}$) depends only on $\omega\textsubscript{s}$ and $\omega\textsubscript{s}$ is inversely proportional to PL\textsubscript{37}\textsuperscript{3/7}, 
the $\dot{P}$ for a star of a given mass and magnetic moment depends only on (PL\textsubscript{37}\textsuperscript{3/7})\textsuperscript{-1}. 
Figure 20 shows that Swift J045106.8-694803 is located in the same region as GX 1+4, SAX J2103.5+4545, 4U 2206+54 and SXP 1062 on a plot of 
log\textsubscript{10}($\dot{P}$) against log\textsubscript{10}(PL\textsubscript{37}\textsuperscript{3/7}).

GX 1+4 is part of a LMXB, it is a slow rotator with a period of about 121 seconds and has been observed to have spun-down by about 2.6 seconds a year \citep{b8}. SAX J2103.5+4545 is part of a HMXB system, 
it is also a slow rotator with a pulse period of 358.166$\pm$0.0005 seconds and a luminosity of (2.0$\pm$0.5)$\times$10\textsuperscript{36} ergs/s. 
It has also been known to spin-up at a rate of 4.69$\pm$0.09 seconds per year \citep{b7}. This means that it has a magnetic field of (1.47$\pm^{0.08}_{0.09}$)$\times$10\textsuperscript{14} Gauss. 4U 2206+54 \citep{b32} 
and SXP 1062 \citep{b35} are also part of HMXB systems, both are known to be spinning down and both were recently identified as having extremely high magnetic fields. 

Highly magnetised neutron stars may be formed if a very magnetic star conserves its magnetic field during 
the neutron stars formation or, if the neutron star is rotating fast enough to start a 
dynamo effect during the first $\sim$20 seconds of its life. This converts heat and rotational energy into magnetic energy and increases the magnetic field \citep{Wickramasinghe}.

The first evidence for highly magnetised accreting neutron stars came from Pizzolato \emph{et al} (2008), who showed that the X-ray source 1E161348-5055 may contain a neutron star with a surface magnetic field 
of $\sim$10\textsuperscript{15}G which is part of a LMXB. This was shortly followed by Bozzo \emph{et al} (2008), who argued that highly magnetised neutron stars may also exist in HMXB. 

\begin{figure*}
 \centering
  \includegraphics[height=170mm,angle=90]{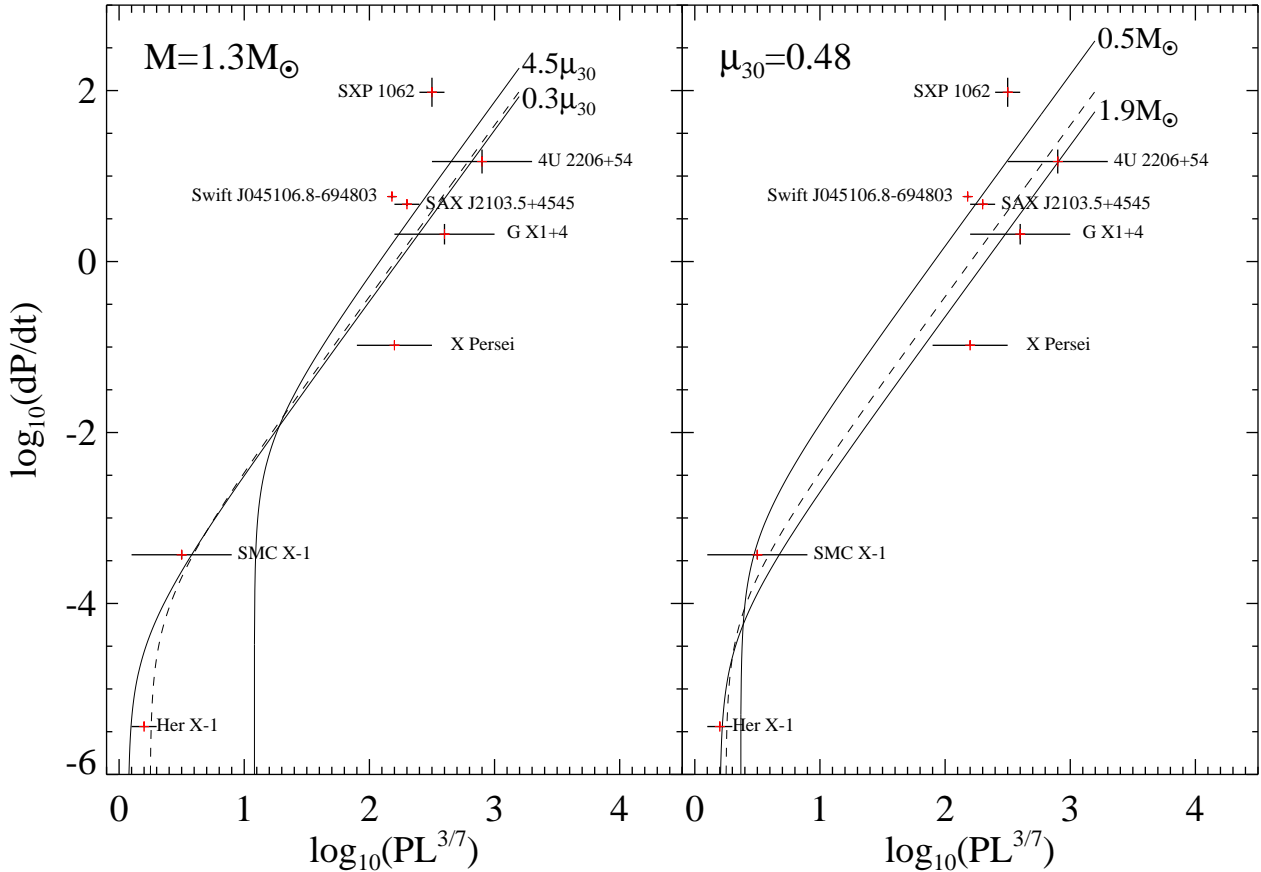}
  \caption{Logarithmic plot of $\dot{P}$ against $PL^{\frac{3}{7}}$ with contours calculated from \citet{b4}. Left panel shows the theoretical curves for three values of 
stellar magnetic moment assuming a neutron star mass of 1.3~M$_\odot$. The dashed line shows the theoretical curve for $\mu_{30}=0.48$. 
The right panel shows the theoretical curves for three values of neutron star mass assuming a magnetic moment of $\mu_{30}=0.48$. The dashed line is the theoretical curve for 
M=1.3~M$_\odot$. Data for Swift J045106.8-694803 and seven other pulsars are included in the plot. 
References are as follows; Her X-1 (Li, Dai and Wang 1995), SMC X-1 (Kahabka and Li 1999), X Per (Delgado-Martí \emph{et al} 2001), GX 1+4 (Ferrigno \emph{et al} 2007 and Chakrabarty and Roche 1997), 
SAXJ2103.5+4545 (Sidoli \emph{et al} 2005), 4U 2206+54 (Reig, Torrejon and Blay 2012) and SXP 1062 (Hénault-Brunet \emph{et al} 2012 and Haberl \emph{et al} 2012).}
\end{figure*}

Our results show that the $\dot{P}$ measured between 2008 and 2011 implies a magnetic field of (1.3$\pm$0.1)$\times$10\textsuperscript{14} Gauss, whilst that measured between 2011 and 2012 
implies a magnetic field of (6$\pm$1)$\times$10\textsuperscript{13} Gauss. The value calculated for the whole duration of observations is (1.2$\pm$0.1)$\times$10\textsuperscript{14} Gauss. 
Taking into account the highest and lowest possible values of B measured 
between 2008 and 2011 and 2011 and 2012, this value becomes  (1.2$\pm^{0.2}_{0.7}$)$\times$10\textsuperscript{14} Gauss. Parameters calculated using equations 1-7 are shown in Table 4.

All values are over the quantum critical value of 4.4$\times$10\textsuperscript{13} Gauss, which means that the physics must be described with quantum field theory rather than classical physics. 
Although Swift J045106.8-694803 is powered by accretion, rather than due to the decay and instabilities of its magnetic field, and hence cannot be referred to as a magnetar yet, it has the potential to 
become one. The Galactic HMXB system containing LSI 61303 recently showed magnetar like behaviour when it underwent two bursts similar to those of soft 
gamma repeaters (SGRs) \citep{torres, papitto}.

\begin{table}
 \centering
\begin{tabular}{@{}cc@{}}
  \hline
  Orbital period&21.631$\pm$0.005 days\\
  Pulse period in October&187.07$\pm$0.04 s\\
  /November 2008&\\
  Pulse period in October 2011&169.8$\pm$0.3 s\\
  Pulse period in July 2012&168.5$\pm$0.2 s\\
  Average Luminosity&(3.4$\pm$0.3)$\times$10$\textsuperscript{36}$ ergs/s\\
   \hline 
  $\dot{P}$&-5.01$\pm$0.06 s/yr\\
            &(-1.59$\pm$0.02)$\times$10\textsuperscript{-7} s/s\\
  T\textsubscript{s}&-34.9$\pm$0.4 yr\\
  log\textsubscript{10}($\dot{P}$)&0.70$\pm$0.01\\
  log\textsubscript{10}(PL\textsubscript{37}\textsuperscript{3/7})&2.04$\pm$0.04\\
  \hline
  $\omega\textsubscript{s}$&0.7$\pm^{0.1}_{0.3}$\\
  n($\omega\textsubscript{s}$)&-3$\pm^{2}_{1}$\\
  $\mu$&(1.2$\pm^{0.2}_{0.7}$)$\times$10\textsuperscript{32} Gauss cm\textsuperscript{-3}\\
  B&(1.2$\pm^{0.2}_{0.7}$)$\times$10\textsuperscript{14} Gauss\\
  \hline
\end{tabular}
\caption{A summary of observed data and results derived using equations 1-7. for Swift J045106.8-694803.}
\end{table}

In conclusion, the spin-up rate of Swift J045106.8-694803 can be explained using Ghosh and Lamb’s (1979) accretion theory provided that it has a magnetic field 
of (1.2$\pm^{0.2}_{0.7}$)$\times$10\textsuperscript{14} Gauss.

\section*{Acknowledgements}
We acknowledge the use of public data from the Swift and RXTE data archive and are grateful for the advice from Gerry Skinner on the Swift/BAT data. 
This paper utilizes public domain data obtained by the MACHO Project, jointly funded by the US Department 
of Energy through the University of California, Lawrence Livermore National Laboratory under contract No. W-7405-Eng-48, by the National Science Foundation through the Center 
for Particle Astrophysics of the University of California under cooperative agreement AST-8809616, and by the Mount Stromlo and Siding Spring Observatory, part of the 
Australian National University. 
The OGLE project has received funding from the European Research Council under the European Community’s Seventh Framework Programme 
(FP7/2007-2013)/ERC grant agreement no. 246678 to AU. 
Optical observations were also made with ESO Telescopes at the La Silla Paranal Observatory under programme ID [088.D-0352(A)] and the SAAO 1.9 meter telescope in South Africa. 
We also thank STFC whose studentships funded HK and ESB.

\appendix

\bsp

\label{lastpage}


\begin{thebibliography}{99}
\bibitem[\protect\citeauthoryear{Beardmore \emph{et al}}{2009}]{b19} Beardmore, A.P., Coe, M.J., Markwadt, C., Osborne, J.P., Baumgartner, W.H., Tueller, J. and Gehrels, N., 2009, The Astronomer's Telegram, \#1901
\bibitem[\protect\citeauthoryear{Bartlett \emph{et al}}{2012, in prep}]{Bartlett} Bartlett, E.S. \emph{et al}, 2012, in prep
\bibitem[\protect\citeauthoryear{Bird \emph{et al}}{2012}]{b34} Bird, A. J., Coe, M. J., McBride, V. A. and Udalski A., 2012, Monthly Notices of the Royal Astronomical Society, 423, 3663
\bibitem[\protect\citeauthoryear{Bonanos \emph{et al}}{2011}]{b9} Bonanos, A.Z., Castro, N., Macri, L.M. and Kudritzki, R-P., 2011, The Astrophysical Journal Letters, 729, L9
\bibitem[\protect\citeauthoryear{Borkowski, Hendrick and Reynolds}{2006}]{b33} Borkowski, K.J., Hendrick, S.P. and Reynolds, S.P., 2006, The Astrophysical Journal, 652, 1259
\bibitem[\protect\citeauthoryear{Bozzo, Falanga and Stella}{2008}]{bozzo} Bozzo, E., Falanga, M. and Stella, L., 2008, The Astrophysical Journal, 683, 1031
\bibitem[\protect\citeauthoryear{Chakrabarty \emph{et al}}{1997}]{b8} Chakrabarty, D., Bildsten, L., Finger, M.H., Grunsfeld, J.M., Koh, D.T., Nelson, R.W., Prince, T.A., Vaughan, B.A. and Wilson, R.B., 1997, The Astrophysical Journal, 481, L101
\bibitem[\protect\citeauthoryear{Chakrabarty and Roche}{1997}]{lgxref} Chakrabarty, D. and Roche, P., 1997, The Astrophysical Journal, 489, 254
\bibitem[\protect\citeauthoryear{Coleiro and Chaty}{2011}]{col} Coleiro, A. and Chaty, S., 2011, Astronomical Society of the Pacific Conference Proceedings, 447, 265
\bibitem[\protect\citeauthoryear{Dachs, Hummel and Hanuschik}{1992}]{b4} Dachs, J., Hummel, W. and Hanuschik, R.W., 1992, Astronomy and Astrophysics Supplement Series, 95, 437
\bibitem[\protect\citeauthoryear{Delgado-Martí \emph{et al}}{2001}]{xperref} Delgado-Martí , H., Levine , A.M., Pfahl, E. and Rappaport, S.A., 2001, The Astrophysical Journal, 546, 455
\bibitem[\protect\citeauthoryear{Dickey and Lockman}{1990}]{b21} Dickey J.M., and Lockman, F.J., 1990, Annual Review of Astronomy and Astrophysics, 28, 215 
\bibitem[\protect\citeauthoryear{Evans \emph{et al}}{2004}]{Evans04} Evans, C.~J., Howarth, I.~D., Irwin, M.~J., Burnley, A.~W. and Harries, T.~J., 2004, Monthly Notices of the Royal Astronomical Society, 353, 601
\bibitem[\protect\citeauthoryear{Evans \emph{et al}}{2006}]{Evans07} Evans, C.~J., Lennon, D.~J., Smartt, S.~J. and Trundle, C., 2006, Astronomy and Astrophysics, 456, 623
\bibitem[\protect\citeauthoryear{Ferrario and Wickramasinghe}{2006}]{Wickramasinghe} Ferrario, L. and Wickramasinghe, D., 2006, Monthly Notices of the Royal Astronomical Society, 367, 1323
\bibitem[\protect\citeauthoryear{Ferrigno \emph{et al}}{2007}]{pandpdotgxref} Ferrigno, C., Segreto, A., Santangelo, A., Wilms, J., Kreykenbohm, I., Denis, M. and Staubert, R., 2007, Astronomy and Astrophysics, 462, 995
\bibitem[\protect\citeauthoryear{Gardiner and Noguchi}{1996}]{b14} Gardiner L. T. and Noguchi M., 1996, Monthly Notices of the Royal Astronomical Society, 278, 191
\bibitem[\protect\citeauthoryear{Ghosh and Lamb}{1979}]{b24} Ghosh, P. and Lamb, F. K., 1979, Astrophysical Journal, 234, 296
\bibitem[\protect\citeauthoryear{Grebenev, Lutovinov and Tsygankov}{2012}]{b23} Grebenev, S.A., Lutovinov, A.A., Tsygankov, S.S. and Mereminskiy, I.A., 2012, arXiv:1207.1750v1
\bibitem[\protect\citeauthoryear{Grimm, Gilfanov and Sunyaev}{2003}]{b26} Grimm, H.J., Gilfanov, M. and Sunyaev, R., Monthly Notices of the Royal Astronomical Society, 339, 793
\bibitem[\protect\citeauthoryear{G{\"u}ver and {\"O}zel}{2009}]{Guver09} G{\"u}ver, T. \& {\"O}zel, F.\ 2009, Monthly Notices of the Royal Astronomical Society, 400, 2050
\bibitem[\protect\citeauthoryear{Haberl \emph{et al}}{2012}]{pdotSXP1062ref} Haberl, F., Sturm, R., Filipović, M. D., Pietsch, W., \& Crawford, E. J., 2012, Astronomy and Astrophysics, 537, 1
\bibitem[\protect\citeauthoryear{Hénault-Brunet \emph{et al}}{2012}]{plSXP1062ref} Hénault-Brunet, V., Oskinova, L. M., Guerrero, M. A., Sun, W., Chu, Y.-H., Evans, C. J., Gallagher III, J. S., Gruendl, R. A. and Reyes-Iturbide, J., 2012, Monthly Notices of the Royal Astronomical Society, 420, L13
\bibitem[\protect\citeauthoryear{Inam and Baykal}{2000}]{b31} Inam, S.C. and Baykal, A., 2000, Astronomy and Astrophysics, 353, 617
\bibitem[\protect\citeauthoryear{Kahabka and Li}{1999}]{smcx1ref} Kahabka, P. and Li, X.-D., 1999, Astronomy and Astrophysics, 345, 117
\bibitem[\protect\citeauthoryear{Lennon}{1997}]{Lennon97} Lennon, D.~J., 1997, Astronomy and Astrophysics, 317, 871 
\bibitem[\protect\citeauthoryear{Li, Dai and Wang}{1995}]{herx1ref} Li, X.-D., Dai, Z.-G., Wang, Z.-R., 1995, Astronomy and Astrophysics, 303, L1
\bibitem[\protect\citeauthoryear{Liu, van Paradijs and van den Hauvel}{2005}]{b28} Liu, Q.Z., van Paradijs, J. and van den Hauvel, E.P.J., 2005, Astronomy and Astrophysics, 442, 1135
\bibitem[\protect\citeauthoryear{Massey}{2002}]{b20} Massey, P., 2002, The Astrophysical Journal Supplement Series, 141, 81
\bibitem[\protect\citeauthoryear{McBride \emph{et al}}{2010}]{b15} McBride, V.A., Bird, A.J., Coe, M.J., Townsend, L.J., Corbet, R.H.D and Haberl, F., 2010, Monthly Notices of the Royal Astronomical Society, 403, 709
\bibitem[\protect\citeauthoryear{Morgan \emph{et al}}{1943}]{MKK1943} Morgan, W.~W., Keenan, P.~C. and Kellman, E.\ 1943, Chicago, Ill., The University of Chicago press [1943]
\bibitem[\protect\citeauthoryear{Negueruela and Coe}{2002}]{b27} Negueruela, I. and Coe, M.J., 2002, Astronomy and Astrophysics, 385, 517
\bibitem[\protect\citeauthoryear{Papitto, Torres and Rea}{2012}]{papitto} Papitto, A., Torres, D.F. and Rea, N., 2012, The Astrophysical Journal, 756, 188 
\bibitem[\protect\citeauthoryear{Paturel \emph{et al}}{2002}]{Paturel02} Paturel, G., Dubois, P., Petit, C., \& Woelfel, F.\ 2002, LEDA, 0 (2002), 0 
\bibitem[\protect\citeauthoryear{Pizzolato \emph{et al}}{2008}]{luca} Pizzolato, F., Colpi, M., De Luca, A., Mereghetti, S., Tiengo, A., 2008, The Astrophysical Journal, 681, 530
\bibitem[\protect\citeauthoryear{Reig}{2011}]{b1} Reig, P., 2011, Astrophysics and Space Science, 332, 1
\bibitem[\protect\citeauthoryear{Reig, Torrejon and Blay}{2012}]{b32} Reig, P., Torrejon, J.M. and Blay, P., 2012, Monthly Notices of the Royal Astronomical Society, 425, 595
\bibitem[\protect\citeauthoryear{Sidoli \emph{et al}}{2005}]{b7} Sidoli, L., Mereghetti, S. and Larsson, S. \emph{et al}, 2005, Astronomy and Astrophysics, 440, 1033
\bibitem[\protect\citeauthoryear{Sturm \emph{et al}}{2012}]{b18} Sturm, R., Haberl, F., Rau, A., Bartlett, E.S., Zhang, X.L., Schady, P., Pietsch, P.W., Greiner, W.J., Coe, M.J. and Udalski, A., 2012, Astronomy and Astrophysics, 542, A109
\bibitem[\protect\citeauthoryear{Torres \emph{et al}}{2012}]{torres} Torres, D.F., Rea, N., Esposito, P., Li, J., Chen, Y. amd Zhang, S., 2012, The Astrophysical Journal, 744, 106 
\bibitem[\protect\citeauthoryear{Turolla and Popov}{2012}]{b35} Turolla, R. and Popov, S.B., 2012, Monthly Notices of the Royal Astronomical Society, 421, L127
\bibitem[\protect\citeauthoryear{Walborn and Fitzpatrick}{1990}]{Walborn90} Walborn, N.~R., \& Fitzpatrick, E.~L.\ 1990, Publications of the Astronomical Society of the Pacific, 102, 379
\bibitem[\protect\citeauthoryear{Wegner}{2006}]{Wegner06} Wegner, W., 2006, Monthly Notices of the Royal Astronomical Society, 371, 185
\bibitem[\protect\citeauthoryear{Wilms, Allen and McCray}{2000}]{b22} Wilms, J., Allen, A. and McCray, R., 2000, The Astrophysical Journal, 542, 914
\bibitem[\protect\citeauthoryear{Zaritsky \emph{et al}}{2004}]{Zaritsky04} Zaritsky, D., Harris, J., Thompson, I.~B. and Grebel, E.~K., 2004, The Astrophysical Journal, 128, 1606 
\end{thebibliography}
\end{document}